\documentclass[aps,pre,twocolumn,showpacs]{revtex4}
\usepackage{epsfig}
\usepackage{amsmath}
\usepackage{times}

\begin{document}

\title{Efficient routing strategies in scale-free networks with limited bandwidth}

\author{Ming Tang$^{1,2}$}
\email{tangminghuang521@hotmail.com}
\author{Tao Zhou$^{1,3}$}
\email{zhutou@ustc.edu}
\affiliation{$^1$Web
Sciences Center, University of Electronic Science and Technology
of China, Chengdu 610054, People's Republic of China\\
$^2$Computer Experimental Teaching Center, University of
Electronic Science and
Technology of China, Chengdu 610054, People's Republic of China\\
$^3$Department of Modern Physics, University of Science and
Technology of China, Hefei 230026, People's Republic of China}
\date{\today}

\begin{abstract}
We study the traffic dynamics in complex networks where each link is
assigned a limited and identical bandwidth. Although the
first-in-first-out (FIFO) queuing rule is widely applied in the
routing protocol of information packets, here we argue that if we
drop this rule, the overall throughput of the network can be
remarkably enhanced. We proposed some efficient routing strategies
that do not strictly obey the FIFO rule. Comparing with the routine
shortest path strategy, the throughput for both Barab\'asi-Albert
(BA) networks and the real Internet, the throughput can be improved
more than five times. We calculate the theoretical limitation of the
throughput. In BA networks, our proposed strategy can achieve 88\%
of the theoretical optimum, yet for the real Internet, it is about
12\%, implying that we have a huge space to further improve the
routing strategy for the real Internet. Finally we discuss possibly
promising ways to design more efficient routing strategies for the
Internet.

\end{abstract}

 \maketitle

\section{Introduction}
Many large-scale traffic networks, such as the Internet, phone
call networks and airport networks, are known to be scale-free
\cite{Albert:2002,Caldarelli2007}. A crucial problem is how to
enhance the transportation capacity, where three kinds of
techniques are usually applied: to design a better assignment of
capacity distribution, to optimize network structure, and to
improve the routing strategy
\cite{Wang2007,Zhou2007,Tadic:2007,Tadic:2009}. Considering the
most widely used routing strategy, the so-called shortest path
(SP) strategy, where packets are sent via the path with the
minimum number of intermediate nodes from the source to the
destination. In a network with heterogeneous degree distribution,
the congestion firstly happens on the hub nodes (they are usually
of the highest loads/betweennesses \cite{Guim:2002}) and soon
spreads to the whole network. Therefore, assigning higher
capacities to the nodes with higher loads will sharply enhance the
throughput of the whole network \cite{Zhao:2005,Liu:2006}. Given
the capacity of each node as well as the SP routing strategy, the
network throughput can be largely enhanced by optimizing the
network structure by using the simulated annealing algorithm
\cite{Danila:2006,Danila:2007}, or by simply removing edges
connecting large-degree nodes \cite{Zhe:2007} or with high edge
betweennesses \cite{Zhang:2007}.

In spite of the effectiveness, to enhance the capacity or to
change the network structure is usually very costly or not
allowed, and thus more efforts have been paid on improving the
routing strategy. Yan \emph{et al.} \cite{Yan:2006} proposed a
highly efficient routing strategy that can automatically detour
the hub nodes, which can remarkably improve the network throughput
without any increasing of computational complexity and can be
further applied in local routing \cite{Wang:2006a}. Sreenivasan
\emph{et al.} introduced a hub avoidance protocol that works
particularly well when the packet-generating rate is close to the
limitation \cite{Sree:2007}. Wang \emph{et al.} \cite{Wang:2006b}
and Kujawski \emph{et al.} \cite{Kujawski:2006} designed the
dynamical routing strategies. Systems with limited queuing length
were also considered
\cite{Echenique:2004,Echenique:2005,Huan:2007,Wu:2008,Wang:2009,Ling:2010,Tang:2009}.

Previous studies overwhelmingly focused on the capacities and/or
limited queuing lengths of nodes, yet paid less attention to the
bandwidths of edges, such as the link capacity of information
packets in the Internet and the number of available seats in the
air transportation networks. Fekete \emph{et al.} showed that much
better performance can be achieved when capacities are distributed
proportional to the expected load of edges \cite{Fekete:2006}. Hu
\emph{et al.} studied the effects of bandwidth on the traffic
capacity of scale-free networks \cite{Hu:2007}. Danila \emph{et
al.} \cite{Danila:2009} proposed an algorithm to minimize the
maximum ratio of edge betweenness to bandwidth. All the
above-mentioned methods embed the first-in-first-out (FIFO)
queuing rule, in contrast, we show in this paper that the FIFO
rule is not necessary and routing strategies without FIFO rule can
remarkably enhance the network throughput and reduce the average
delivering time. Simulation results on artificially generated
scale-free networks as well as real Internet demonstrate the
advantages of out proposed strategy.

\section{Model}

In our model, all nodes are treated as both hosts and routers for
generating and delivering packets and each link has the same
maximum capacity of delivering packets. For simplicity, we set the
capacity of each link~(i.e., bandwidth)~$B=1$, namely only one
packet can be delivered via a link at each time step. Thus, at
each time step a node $i$ with $k_i$ links can deliver at most
$k_i$ packets one step toward their destinations. The transport
processes is as follows.

(1) At each time step, $\lambda N$ packets are generated with
randomly chosen starting points and destinations, where $N$ is the
number of nodes. Each newly created packet is placed at the end of
the queue of its starting node.

(2) For each node, according to the routing strategy (the details
will be introduced in the next section), the packets are checked
successively on a first-in basis. Once the link suggested by the
routing strategy is free, the packet will be accordingly delivered
to its neighboring node, otherwise, it will be stay in the queue.
Therefore, although packets are checked on a first-in basis, they
do not strictly follow the FIFO rule.

(3) Packet arriving at its destination is removed from the system,
otherwise is queued up.

Considering the order parameter~\cite{Arenas:2001}
\begin{equation}\label{eq:slope}
\eta = \frac{1}{\lambda N}\lim_{t\rightarrow \infty} \frac{\langle
S(t+\Delta t) - S(t)\rangle}{\Delta t},
\end{equation}
where $S(t)$ denotes the total number of packets at time step $t$.
When the system is in the congested phase, $\eta>0$. The threshold
$\lambda_c$ separates balance phase ($\eta=0$) and the congested
phase ($\eta>0$), which is the most significant quantity for
transportation networks: The larger, the better.

\section{Routing Strategies}

When simply applying the SP strategy, packets are more likely to
pass through the links with high betweenness, which may lead to
congestion on these links.  Therefore, to enhance the congestion
threshold $\lambda_c$, a routing strategy should adequately
utilize the links with low betweenness. From this point, we
propose some more efficient routing strategies as follows.

\begin{figure}
\epsfig{figure=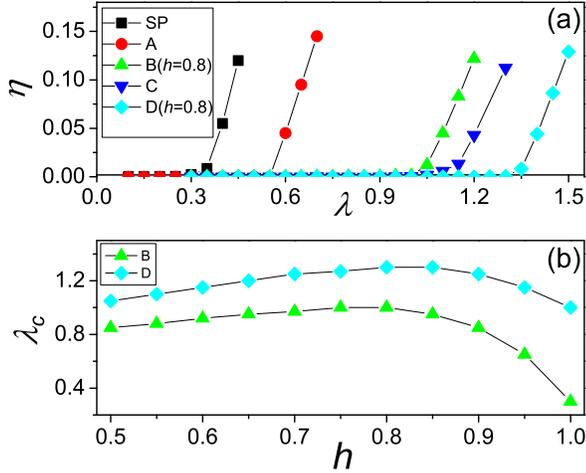,width=1.0\linewidth}\vspace{-0.5cm}
\caption{(color online). Comparison among different routing
strategies in BA networks with $N=2000$ and $m=m_0=3$, where $m_0$
and $m$ are numbers of starting nodes and new links added at every
time step, respectively. (a) The order parameter $\eta$ as a
function of $\lambda$. (b) The network throughput $\lambda_c$ as a
function of the traffic-awareness parameter.} \label{eta1}
\end{figure}

A. The FIFO queuing discipline is followed strictly, namely each
node $i$ delivers the packets one by one from the foremost to the
last (of course, it can deliver at most $k_i$ packets since
$B=1$), and among all the unoccupied links, each packet will
choose the link along the shortest path to the destination. If
there are several links satisfying the requirement, one of them is
randomly selected. Notice that, a packet may detour if all the
unoccupied links point to nodes who is further to the destination
than the current node.

B. At the beginning of each time step, we set a time delay
$\tau_{i\ell}=\tau_{\ell i}=0$ on every link. Different from the
strict FIFO rule, each node $i$ checks packets one by one
following FIFO rule yet may not deliver them in such order. A
packet will choose a neighboring node $\ell$ towards its
destination $j$ with the smallest value of \emph{effective
distance} denoted by
\begin{equation}\label{eq:strategy B}
d_{B}(\ell)=hd_{\ell j} + (1-h)\tau_{i\ell},
\end{equation}
where $d_{\ell j}$ is the topological distance between nodes
$\ell$ and $j$, and $h$ is the traffic-awareness parameter. If the
link $(i\rightarrow \ell)$ is unoccupied (i.e., $\tau_{i
\ell}=0$), the packet will be delivered, otherwise this packet
will not be delivered in this time step but still queued up in its
current position. Whatever this packet has been delivered, we set
$\tau_{i \ell}\leftarrow \tau_{i \ell}+1$. In this way, packets in
the later position have the chanced to be delivered at an earlier
time, and they are aware of the approximated waiting time of each
candidate link and thus may choose a link who points further node
to the destination but is not congested. In contrast, packets
willing to go through central links may be delayed even they lie
in top positions of the queue.

C. It is known that the betweenness centrality of a link
$(i\leftrightarrow\ell)$ is strongly correlated with its product
degree $k_ik_\ell$~\cite{Holme:2002}. Accordingly, we assign a
weight to every link as
\begin{equation}\label{eq:strategy C}
w_{i\ell}=(k_{i}k_{\ell})^{\theta},
\end{equation}
where $\theta$ is an adjustable parameter. Similar to the strategy
A, this strategy strictly obeys the FIFO rule but it uses weighted
shortest path.

D. This strategy is a weighted version of the strategy B, also
does not obey the strict FIFO rule. The Eq. (\ref{eq:strategy B})
is replaced by a weighted version according to Eq.
(\ref{eq:strategy C}), as
\begin{equation}\label{eq:strategy D}
d_{D}(\ell)=hd_{\ell j} + (1-h)\tau_{i\ell}w_{i\ell}.
\end{equation}

\begin{figure}
\epsfig{figure=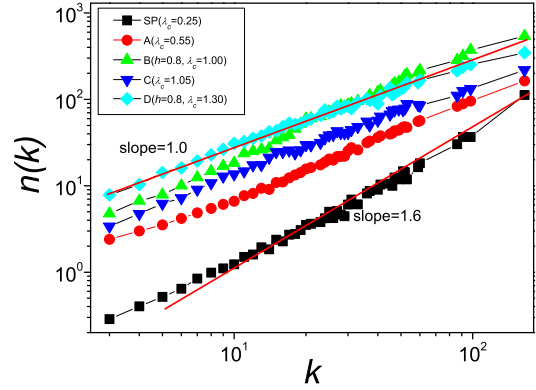,width=1.0\linewidth}\vspace{-0.5cm}
\caption{(color online). The average number of packets $n(k)$ over
the nodes with degree $k$. Network parameters are the same to
those of figure 1.} \label{nk1}
\end{figure}

\section{Simulation Results}

This section compares different routing strategies in
Barabasi-Albert (BA) networks~\cite{BA:1999}, where the
performance is quantified by the network throughput $\lambda_c$:
the larger the better. The parameter $\theta$ is fixed as $\theta
=0.25$ since at that point the weighted betweenness of a node is
approximately linearly correlated with its degree. As shown in
Fig. 1(b), subject to the largest $\lambda_c$, the optimal $h$ for
both strategies B and D is about 0.8, and thus in the simulation,
$h=0.8$ is also fixed.

Fig.~\ref{eta1}(a) reports the phase transition for different
routing strategies, where $\lambda_c = 0.25 (SP) < 0.55 (A) < 1.00
(B, h=0.8) < 1.05 (C) < 1.30 (D, h=0.8)$. The SP strategy is the
worst one since it cannot well utilize the capacities of
small-betweenness links. Under the SP strategy, too many packets
jam at the large-degree nodes and the number of packets queuing up
at a node is superlinearly correlated with its degree. As shown in
Fig. 2, $B(k) \sim n(k) \sim k^{\alpha}$ with $\alpha \approx
1.6$, where $B(k)$ is the average betweenness over nodes with
degree $k$ and $n(k)$ is the average number of packets over nodes
with degree $k$. This result is in accordance with previous
observations~\cite{Goh:2001, Barthelemy:2003}. Much differently,
for the proposed strategies (A-D), the small-betweenness links are
well utilized and thus the number of packets waiting at a node is
more or less linearly correlated with its degree (see also Fig.
2).

Although strategy A makes all links almost fully utilized, massive
bandwidths (links) are squandered since many packets will detour and
pass long paths to the destinations. Taking into account both the
shortest path to the destination and the time delay of a candidate
link, strategy B introduces the effective distance and thus a vacant
path may be selected instead of the shortest path. Compared with the
strict FIFO queuing discipline in strategy A, the strategy B is more
flexible and performs better. As shown in Fig.~\ref{nk1}, the slope
of the $n(k)$ curve for strategy B ($h=0.8$) is much less than that
for strategy A, indicating that the small-betweenness links are
utilized more effectively.

Strategy C makes the real traffic load of each link more or less
the same, which has a slightly higher threshold than that of the
strategy B. However, this strategy isn't optimal due to the
time-dependant fluctuation of the number of packets passing
through each link~\cite{Cai2007,Kujawski:2007}. That is to say,
the strategy C is very good as a static strategy while it fails to
capture the real-time traffic in the network. Once the effective
weighted distance is introduced into the strategy D (in the same
way to what we did for the strategy B), $\lambda_c$ will increase
to $1.30$.

In the ideal condition where $\sum_{i}k_i$ packets are delivered and
each of them takes the shortest path to the destination without any
delay. The theoretically largest throughput reads
\begin{equation}\label{eq:lambdac}
\lambda_u=\frac{\sum_{i}k_i}{\langle L \rangle N}=\frac{\langle
k\rangle}{\langle L\rangle}.
\end{equation}
In BA networks with $N=2000$, $\langle k\rangle = 6$ and $\langle
L \rangle \approx \langle d \rangle = 4.0589$, we have $\lambda_u
= 6/4.0589 \approx 1.48$. Of course, owing to the complicated
local structure of networks and the real-time fluctuations,
actually, the theoretical limitation $\lambda_u$ can't be
achieved. Not so bad, the throughput of the strategy D
($\lambda_c=1.30$) is about 88\% of the theoretical limitation.

\begin{figure}
\epsfig{figure=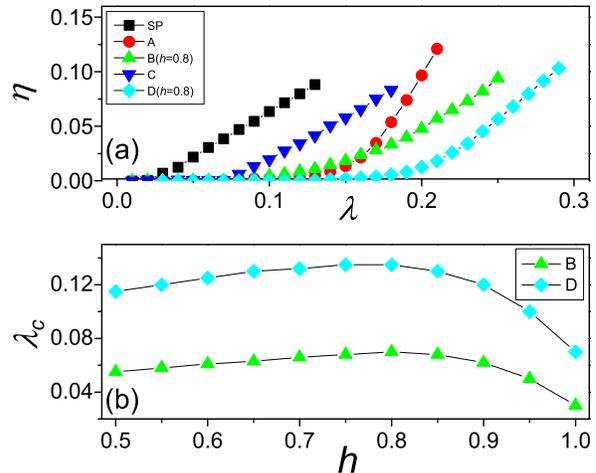,width=1.0\linewidth}\vspace{-0.5cm}
\caption{(color online). Comparison among different routing
strategies for the real Internet. (a) The order parameter $\eta$ as
a function of $\lambda$. (b) The network throughput $\lambda_c$ as a
function of the traffic-awareness parameter.} \label{eta2}
\end{figure}

\begin{figure}
\epsfig{figure=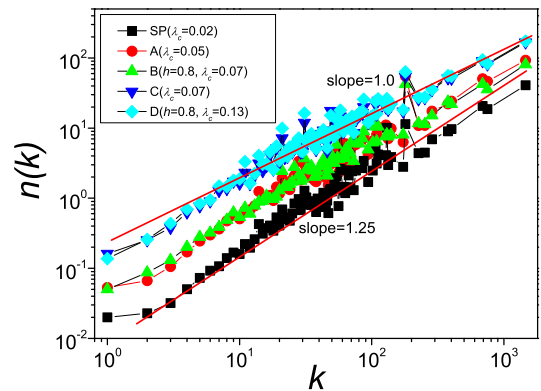,width=1.0\linewidth}\vspace{-0.5cm}
\caption{(color online). The average number of packets $n(k)$ over
the nodes with degree $k$ for the real Internet.} \label{nk2}
\end{figure}

\section{Analyzing Real Internet}
In this section, we will apply the proposed strategies on the real
Internet at autonomous system (AS) level~\cite{Internet:1999}, where
the network size $N=6474$, the average degree $\langle
k\rangle\approx3.88$, the average distance $\langle
d\rangle\approx3.71$, the maximum degree $k_{max}=1458$ and the
power-law exponent of the degree distribution $\gamma=2.2\pm0.1$. As
shown in Fig.~\ref{eta2}(a), the proposed routing strategies are
more efficient than the simples SP strategy or the strict FIFO rule
of the strategy A: $\lambda_c = 0.02 (SP) < 0.05 (A) < 0.07 (B,
h=0.8) = 0.07 (C) < 0.13 (D, h=0.8)$. Similar to the observations
for BA networks, the strategy D ($h=0.8$) also performs the best,
with the corresponding $n(k)$ curve being of slope about 1 (see
Fig.~\ref{nk2}). From Fig.~\ref{eta2}(b), we are happy to see that
the optimal $h$ for the strategies B and D is about 0.8, same to the
case for BA networks, indicating that this optimal value may be not
very sensitive to the network structure.

Owing to the structural properties of the real Internet, such as
disassortative mixing, clustering coefficient and community
structure~\cite{Satorras:2001, Vazquez:2002, Zhou:2004, Zhou:2006,
Zhang2008}, as shown in Fig.~\ref{nk2}, there is much greater
fluctuation of mean packet number $n(k)$ compared with the case of
BA networks. It implies that some links and nodes are overload while
some others may be largely wasted. As a result, the throughput of
the strategy D ($\lambda_c=0.13$) is only about 12\% of the
theoretical limitation ($\lambda_u = 3.88/3.71 \approx 1.04$), which
is much less than 88\% in BA scale-free networks. This result to
some extent explains why it is necessary to install interchangeable
paths or increase bandwidths of those links with high
link-betweenness in order to enhance the total capacity of the
Internet~\cite{Serrano:2005}, and it leaves a huge space for us to
further improve the throughput via designing a smart routing
strategy properly taking into account the structural features of the
real Internet.

\section{Conclusion and Discussion}

In conclusion, we have studied the traffic dynamics with limited
link bandwidth. Although the first-in-first-out (FIFO) queuing rule
is applied everywhere, here we argue that if we drop this rule, the
overall throughput of the network can be remarkably enhanced. Taking
the strategy D as an example, compared with the shortest path
strategy (SP) and the strategy A with strict FIFO rule, the
throughput is enhanced to more than five times and more then two
times in BA networks. We have also applied this strategy to the real
Internet, and compared with the SP strategy and the strategy A, the
improvements are 6.5 times and 2.6 times. Another probable advantage
(not yet fully demonstrated) is that the optimal value of the key
parameter $h$ seems not very sensitive to the network structure, as
for the BA networks and the real Internet, the optimal values are
both about 0.8.

In BA networks, the performance of the strategy D is close to the
theoretical limitation (i.e., 88\% of the theoretical optimum).
However, for the real Internet, this fraction becomes much lower,
about 12\%. It indicates that the structural properties of BA
networks are far different from the real Internet, and the
complicated local structure of the Internet, such as mixing
patterns, clustering, cliques, loop structure and community
structure, makes the design of an advanced routing strategy much
harder. The further improvement can be achieved by (i) real-time
routing strategies yet they ask for great computational power and
other advanced techniques; (ii) more smart routing strategy taking
into consideration the structural features of the Internet. Or,
maybe we should follow the suggestions by Zhao \emph{et al.}
\cite{Zhao:2005} and Serrano \emph{et al.} \cite{Serrano:2005} that
the bandwidth of each link should be carefully assigned in a
heterogeneous way.

\acknowledgments This work was supported by the National Natural
Science Foundation of China (Grant Nos. 90924011 and 10635040).

\end{document}